\documentclass[%
 reprint,
superscriptaddress,
 amsmath,amssymb,
 aps,
pra,
]{revtex4-2}
\usepackage[dvipsnames]{xcolor}
\usepackage[dvipsnames]{xcolor}
\usepackage{physics}
\usepackage{amsmath}     
\usepackage{amssymb}     
\usepackage[version=4]{mhchem}

\usepackage{mhchem}      
\usepackage{graphicx}
\usepackage{dcolumn}
\usepackage{bm}

\usepackage[export]{adjustbox}
\begin{document}
\preprint{APS/123-QED}

\title{Tripartite Entanglement Generation in Atom-Coupled Dual Microresonators System}

\author{Abhishek Mandal}%
\email{mandalabhi350@kgpian.iitkgp.ac.in}
\affiliation{%
Department of Physics, IIT Kharagpur, 721302, West Bengal, India}%
\author{Joy Ghosh}
\affiliation{%
School of Nanoscience and Technology, IIT Kharagpur , 721302, West Bengal, India
}%
\author{Maruthi Manoj Brundavanam}%
\affiliation{%
Department of Physics, IIT Kharagpur, 721302, West Bengal, India}%
\author{Shailendra K. Varshney}
 \affiliation{Department of Electronics and Electrical Communication Engineering, IIT Kharagpur, 721302, West Bengal, India.}
 
\begin{abstract}
 In this work, we investigate the emergence and control of genuine tripartite entanglement in a hybrid cavity quantum electrodynamics architecture consisting of two linearly coupled single-mode resonators, one of which interacts coherently with a two-level atom. An analytical framework is developed in weak-driving regime, where the system dynamically supports a delocalized hybrid excitation shared by the two photonic modes and the atomic degree of freedom. Tripartite concurrence fill has been used to characterize and identify parameter regimes of maximal multipartite quantum correlation that can be generated in this model. Additionally, we demonstrate how dissipative rates and detuning asymmetries govern the conversion of bipartite entanglement into a genuinely tripartite state, establishing a controllable transition from localized Jaynes–Cummings correlations to delocalized photonic–atomic entanglement networks. These findings outline a clear route to engineering steady-state multipartite quantum resources in coupled cavity–atom platforms, with direct relevance to quantum networking, distributed quantum information processing, and photonic state routing in scalable quantum architectures.
\end{abstract}

\maketitle

\section{Introduction}
Quantum entanglement is one of the most fundamental features of quantum mechanics and plays a central role in quantum information science. Over the past decades, entanglement has been recognized as the key resource for quantum communication, quantum computation, and quantum metrology. While bipartite entanglement has been extensively characterized and is now routinely generated in various physical platforms, the study of multipartite entanglement has revealed qualitatively new correlations that go beyond pairwise entanglement and are essential for scaling quantum technologies \cite{PhysRevA.62.062314}. Entanglement serves as a cornerstone of quantum information science, underpinning a wide range of technological advancements, including quantum teleportation, quantum key distribution, dense coding, and precision metrology. While the quantification and characterization of entanglement are widely formulated for bipartite systems, recent theoretical and experimental advances have extended these frameworks to encompass multipartite entanglement across various microscopic platforms, including photonic qubits, trapped ions, and atomic ensembles. Tripartite entanglement represents the minimal non-trivial case of the genuine multipartite criterion, exhibiting distinct non-classical correlations such as GHZ and W-type states, that are nonequivalent under stochastic local operations and classical communication. These states possess different structural and operational properties relevant to distributed quantum protocols, quantum error correction, and network-based quantum computation, making the generation, detection, and manipulation of tripartite entanglement a subject of significant research interest. Among multipartite scenarios, tripartite entanglement represents the minimal nontrivial extension, providing a fertile test bed for investigating genuine multipartite correlations. It is well established that three qubits entanglement manifests in two nonequivalent classes: the Greenberger–Horne–Zeilinger (GHZ) class and the W class \cite{PhysRevA.65.032108}. The GHZ states exhibit maximal nonlocal correlations and have been instrumental in foundational tests of quantum mechanics \cite{PhysRevLett.65.1838}, while W states are more robust to local decoherence and particle loss, making them attractive for applications such as quantum communication and error-resilient protocols \cite{PhysRevA.61.052306,RevModPhys.71.S288}. To characterize these states, various entanglement measures have been developed, including the three-tangle \cite{Meyer2001GlobalEI}, global entanglement measures \cite{PhysRevA.68.042307}, geometric entanglement \cite{PhysRevA.72.012337}, and, more recently, concurrence-based multipartite measures such as the concurrence fill. These tools have significantly expanded our ability to quantify and classify genuine tripartite correlations in both pure and mixed states.

Experimentally, multipartite entanglement has been demonstrated in a wide variety of platforms, including trapped ions , cold atoms, optical photons, and superconducting circuits \cite{PhysRevLett.134.050201}. Recent advances in hybrid quantum architectures that combine atomic, photonic, and nonlinear elements have enabled novel routes toward robust generation and control of multipartite entanglement \cite{RevModPhys.86.153}. In particular, coupled-cavity qubit systems have emerged as versatile candidates for engineering entanglement between photonic and atomic modes. The interplay between coherent driving, strong light–matter coupling, and engineered dissipation allows not only the creation of bipartite entanglement but also the stabilization of genuine multipartite entangled states in the steady state \cite{PhysRevLett.88.197901}. These approaches are promising for constructing distributed quantum networks and scalable architectures for quantum computation. Hybrid cavity QED and circuit QED systems offer unique opportunities for realizing such entanglement. By exploiting the Jaynes–Cummings interaction between a two-level system and a resonator mode, along with inter-cavity photon hopping and nonlinearities, it is possible to generate GHZ-like or W-like correlations among three subsystems \cite{PhysRevLett.88.197901}. Moreover, recent works have emphasized the role of dissipation engineering and coherent control in stabilizing multipartite entanglement against environmental noise, thereby enhancing experimental feasibility. Notably, multipartite entanglement in hybrid resonator–atom systems has also been investigated in the context of quantum routing, quantum state transfer \cite{RevModPhys.90.035005}, and quantum metrology \cite{PhysRevLett.96.010401}.In this work, we study a hybrid model comprising two coupled resonators and a two-level atom, driven coherently and subject to dissipation, with the aim of generating robust tripartite entanglement in the steady state. Using both numerical simulations and analytical approaches, we identify parameter regimes where concurrence-based measures, such as concurrence fill, reveal strong tripartite correlations. Our analysis contributes to the broader effort of developing hybrid cavity–qubit platforms as building blocks for distributed quantum information processing and scalable entanglement generation.

\section{Hamiltonian description}

We consider a hybrid quantum system composed of two evanescently coupled optical microresonators and a two-level atom positioned in the near field of one of the resonators, such that the atom interacts with the evanescent tail of the resonator mode. The atom is directly coupled to the first resonator, while both resonators are coherently driven by external pump fields. The overall system dynamics are governed by a Hamiltonian incorporating photon hopping between the resonators, a Jaynes--Cummings--type atom--resonator interaction with coupling strength $g$, and coherent driving terms $\Omega_{1,2}$. 
\begin{figure}
    \centering
    \includegraphics[width=\linewidth]{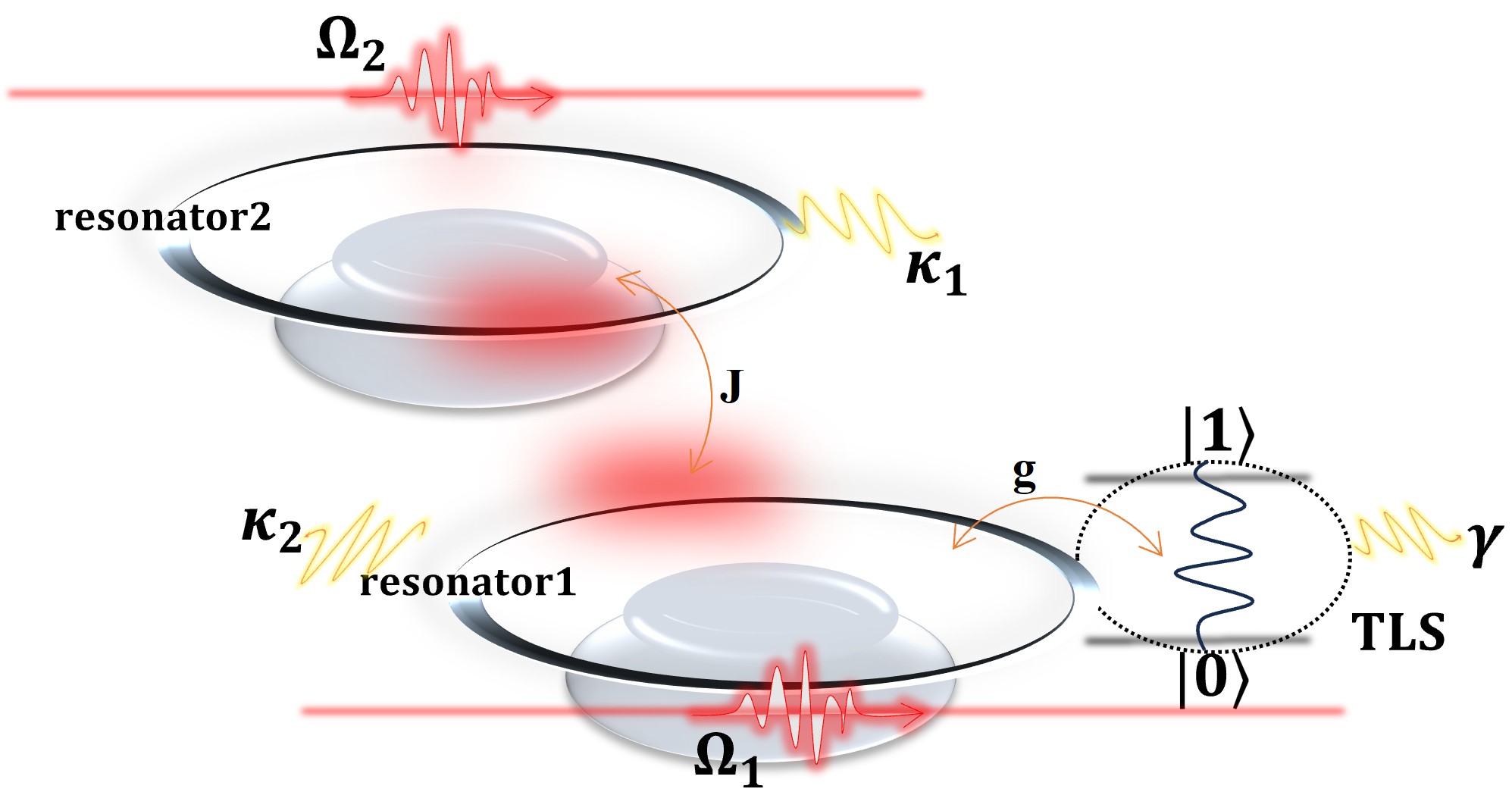}
    \caption{Schematic diagram of the proposed model to generate tripartite entanglement.}
    \label{fig:FIG-1}
\end{figure}
In a realistic experimental implementation \cite{vahala2003optical}, the two-level atom can be realized by several physical systems that effectively behave as artificial or natural two-level emitters. Suitable candidates include alkali atoms such as \textsuperscript{87}Rb or \textsuperscript{133}Cs, where the optical transitions (e.g., the D$_2$ line at 780~nm for \textsuperscript{87}Rb) are perfectly aligned to high-$Q$ silica or silicon nitride microresonator modes\cite{PhysRevLett.91.043902}. Alternatively, semiconductor quantum dots embedded in microdisk or microsphere resonators, or color centers such as nitrogen--vacancy (NV) centers in diamond coupled to whispering-gallery modes, can serve as effective two-level systems with controllable coupling strengths\cite{RevModPhys.91.025005}. The photon hopping rate $J$ between the resonators is determined by their separation and evanescent overlap.
The complete Hamiltonian of the system is given by (taking $\hbar = 1$)
 \begin{eqnarray}
\hat{H} &=& \omega_1 \hat{a}_1^{\dagger} \hat{a}_1 
+ \omega_2 \hat{a}_2^{\dagger} \hat{a}_2 
+ \frac{\Delta_{a}}{2} \sigma_{z} 
+ J\left(\hat{a}_1^{\dagger} \hat{a}_2 + \hat{a}_2^{\dagger} \hat{a}_1 \right) \nonumber\\
&& + g\left(\hat{a}_1^{\dagger} \sigma_{-} + \hat{a}_1 \sigma_{+}\right)
+ \Omega_{1}\left(\hat{a}_1^{\dagger} e^{-i\omega_L t} + \hat{a}_1 e^{i\omega_L t}\right) \nonumber\\
&& + \Omega_{2}\left(\hat{a}_2^{\dagger} e^{-i\omega_L t} + \hat{a}_2 e^{i\omega_L t}\right)
\label{eq1}
\end{eqnarray}

To eliminate the explicit time dependence arising from the driving terms, we consider a rotating reference frame at the laser frequency $\omega_{L}$. This is achieved by applying the unitary transformation $U = \exp\left[i \omega_{L} t \left( \hat{a}_1^{\dagger} \hat{a}_1 + \hat{a}_2^{\dagger} \hat{a}_2 + \frac{\sigma_{z}}{2} \right)\right]$.
This transformation renders the Hamiltonian time-independent, which is essential for steady-state and spectral analysis. Under this operation, the transformed Hamiltonian is given by $\hat{H}' = U \hat{H} U^{\dagger} - i U \frac{dU^{\dagger}}{dt}$.
From this transformation, we introduce the detuning parameters between each resonator mode and the driving laser, defined as $\Delta_{1,2} = \omega_{1,2} - \omega_{L}$.
Accordingly, the Hamiltonian in the rotating frame can be rewritten as
\begin{eqnarray}
\hat{H} &=& \Delta_1 \hat{a}_1^{\dagger} \hat{a}_1 
+ \Delta_2 \hat{a}_2^{\dagger} \hat{a}_2 
+ \frac{\Delta_{a}}{2} \sigma_{z} 
+ J\left(\hat{a}_1^{\dagger} \hat{a}_2 + \hat{a}_2^{\dagger} \hat{a}_1 \right) + g\times\nonumber\\
&& \left(\hat{a}_1^{\dagger} \sigma_{-}  + \hat{a}_1 \sigma_{+}\right)
+ \Omega_{1}\left(\hat{a}_1^{\dagger} + \hat{a}_1\right)
+ \Omega_{2}\left(\hat{a}_2^{\dagger} + \hat{a}_2\right)
\label{eq2}
\end{eqnarray}

The dynamics of the hybrid resonator-atom-resonator system, as described by Eq. (\ref{eq2}), are governed by the Lindblad master equation, which captures both coherent and dissipative processes. It can be expressed as
\begin{eqnarray}
\frac{d\rho}{dt} &=& -i [\hat{\mathcal{H}}, \rho] 
+ \kappa_{1} \, \mathcal{D}[\hat{a}_{1}] \rho
+ \kappa_{2} \, \mathcal{D}[\hat{a}_{2}] \rho
+ \gamma \, \mathcal{D}[\hat{\sigma}_{-}] \rho,
\label{eq:lindblad}
\end{eqnarray}
where $\rho$ is the density matrix of the total system, and $\mathcal{D}[O]\rho$ is the Lindblad superoperator defined as $\mathcal{D}[O]\rho = O \rho O^{\dagger}
- \frac{1}{2} \left( O^{\dagger} O \rho + \rho O^{\dagger} O \right)$,
for a collapse operator $O \in \{\hat{a}_{1}, \hat{a}_{2}, \hat{\sigma}_{-}\}$. The dissipation terms $\kappa_{1,2}$ represent photon loss rates in the two resonators, while $\gamma$ denotes the spontaneous decay rate of the atom. These terms are included in the Lindblad form to account for both cavity photon leakage and atomic relaxation processes. The master Eq. (\ref{eq:lindblad}) is typically solved numerically using solvers such as \textit{mesolve}, which evolve $\rho(t)$ under both coherent (Hamiltonian) and incoherent (dissipative) dynamics.

For simplicity, we consider equal optical decay rates for both resonators, $\kappa_{1} = \kappa_{2} = \kappa$, while the atomic decay rate is denoted by $\gamma$. The corresponding non-Hermitian effective Hamiltonian can then be written as
\begin{eqnarray}
\hat{H}_{\text{eff}} &=& \hat{H}
- \frac{i\kappa}{2} \left( \hat{a}_{1}^{\dagger}\hat{a}_{1} + \hat{a}_{2}^{\dagger}\hat{a}_{2} \right)
- \frac{i\gamma}{2} \, \hat{\sigma}_{+}\hat{\sigma}_{-}.
\label{eq:heff}
\end{eqnarray}

After incorporating these dissipative terms, the full effective Hamiltonian in the rotating frame becomes
\begin{eqnarray}
\hat{H} &=& 
(\Delta_{1} - \tfrac{i\kappa}{2}) \hat{a}_{1}^{\dagger} \hat{a}_{1}
+ (\Delta_{2} - \tfrac{i\kappa}{2}) \hat{a}_{2}^{\dagger} \hat{a}_{2}
+ \left( \tfrac{\Delta_{a}}{2} - \tfrac{i\gamma}{2} \right) \hat{\sigma}_{z} \nonumber\\
&& + J \left( \hat{a}_{1}^{\dagger} \hat{a}_{2} + \hat{a}_{2}^{\dagger} \hat{a}_{1} \right)
+ g \left( \hat{a}_{1}^{\dagger} \hat{\sigma}_{-} + \hat{a}_{1} \hat{\sigma}_{+} \right) \nonumber\\
&& + \Omega_{1} \left( \hat{a}_{1}^{\dagger} + \hat{a}_{1} \right)
+ \Omega_{2} \left( \hat{a}_{2}^{\dagger} + \hat{a}_{2} \right).
\label{eq5}
\end{eqnarray}

\section{Entanglement formulation: bipartite and tripartite}

Formally, entangled particles share a joint quantum state, represented by a non-separable wavefunction, where measurements on one particle instantly determine the corresponding properties of the other(s)\cite{RevModPhys.81.865}. For bipartite systems, several robust measures exist to characterize quantum correlations, including the von Neumann entropy of the reduced density matrix for pure states\cite{PhysRevA.54.3824}, the concurrence and entanglement of formation for mixed two-qubit systems\cite{PhysRevLett.80.2245,PhysRevA.53.2046}, and the negativity or logarithmic negativity, which are based on the Peres–Horodecki partial transpose criterion\cite{PhysRevA.65.032314,HORODECKI19961,PhysRevLett.77.1413}. These measures effectively capture entanglement between two systems, as they quantify the non-classical correlations that cannot be simulated by local operations and classical communication (LOCC)\cite{Nielsen_Chuang_2010}.

The more challenging scenario is the quantification of entanglement in multipartite systems. In this context, the concurrence fill \cite{PhysRevLett.114.140402} serves as a measure of genuine tripartite entanglement, which is evaluated from the reduced density matrices corresponding to different bipartitions of the full system. For each bipartite partition, the concurrence is computed using the well-known Wootters formula\cite{PhysRevLett.80.2245}
\begin{equation}
\mathcal{C}(\rho) = \max\left(0, \lambda_1 - \lambda_2 - \lambda_3 - \lambda_4 \right),
\end{equation}
where $\{\lambda_j\}$ are the square roots of the eigenvalues, arranged in decreasing order, of the non-Hermitian matrix 
$\rho (\sigma_y \otimes \sigma_y) \rho^{*} (\sigma_y \otimes \sigma_y)$.
This formulation is valid when the reduced density matrices effectively correspond to two-level subsystems, which is ensured by truncating each resonator's Hilbert space to the lowest two Fock states. 

To quantify genuine tripartite entanglement, Xie et al. \textbf{cite} proposed a geometric measure based on the entanglement polygon inequality, which governs the distribution of bipartite entanglement across three subsystems. This is expressed as:
\begin{equation}
C^{2}_{i(jk)} \leq C^{2}_{j(ki)} + C^{2}_{k(ij)},
\end{equation}
where $C_{i(jk)}$ represents the bipartite concurrence between the subsystem $i$ and the combined subsystem of $j$ and $k$. The concurrence fill, a measure of genuine tripartite entanglement, quantifies correlations beyond pairwise entanglement and is visualized as the area of a concurrence triangle (see FIG 2). It is defined as\cite{PhysRevLett.127.040403}:
\begin{eqnarray}
\mathcal{F}_{123} &\equiv& \left[ \frac{16}{3}
\left(Q - C^2_{1(23)}\right)
\left(Q - C^2_{2(13)}\right)
\left(Q - C^2_{3(12)}\right) Q \right]^{1/4}\label{eq:8}
\end{eqnarray}
where $Q = \frac{1}{2} \left( C^2_{1(23)} + C^2_{2(13)} + C^2_{3(12)} \right)$ ensures entanglement across all pairwise subsystems. The mathematical identity, nonzero $\mathcal{C}_{\text{fill}}$ confirms the presence of genuine tripartite entanglement, excluding biseparability and verifying inseparability across all bipartitions.

\subsection{Resonator-Atom-Resonator Tripartite Model}

Fig.\ref{fig:FIG-1} illustrates the tripartite concurrence structure, where the vertices represent subsystems: resonator 1, an atom, and resonator 2. Each edge of the triangle corresponds to the squared concurrence between one subsystem and the composite of the remaining two. Under weak driving conditions, the quantum state of the system is described by a pure state of the form:
\begin{eqnarray}
|\psi\rangle &=& c_0 |000\rangle + c_1 |100\rangle + c_2 |010\rangle + c_3 |001\rangle,
\label{eq:state}
\end{eqnarray}
where $c_0, c_1, c_2, c_3$ are complex amplitudes satisfying the normalization condition:
\begin{eqnarray}
|c_0|^2 + |c_1|^2 + |c_2|^2 + |c_3|^2 = 1.
\label{eq:normalization}
\end{eqnarray}
The density matrix for the pure state is given by:
\begin{eqnarray}
\rho &=& |\psi\rangle \langle \psi|.
\label{eq:density_matrix}
\end{eqnarray}
In the ordered basis $\{|000\rangle, |100\rangle, |010\rangle, |001\rangle\}$, the density matrix is an $4\times 4$ outer product, but only elements surviving partial traces are needed for reduced density matrices. The reduced density matrix for subsystem $i$ (e.g., resonator 1) is obtained via:
\begin{eqnarray}
(\rho_i)_{mn} &=& \sum_{y,z} \langle m,y,z|\rho|n,y,z\rangle, \quad m,n \in \{0,1\}.
\label{eq:reduced_density_matrix}
\end{eqnarray}
For subsystem 1, the matrix elements are:
\begin{eqnarray}
(\rho_1)_{00} &=& |c_0|^2 + |c_2|^2 + |c_3|^2, \\
(\rho_1)_{11} &=& |c_1|^2, \\
(\rho_1)_{01} &=& \sum_{y,z} \langle 0,y,z|\rho|1,y,z\rangle = c_0 c_1^*, \\
(\rho_1)_{10} &=& (\rho_1)_{01}^* = c_1 c_0^*.
\end{eqnarray}
Thus, the reduced density matrix for subsystem 1 is:
\begin{eqnarray}
\rho_1 &=&
\begin{pmatrix}
|c_0|^2 + |c_2|^2 + |c_3|^2 & c_0 c_1^* \\
c_1 c_0^* & |c_1|^2
\end{pmatrix}.
\end{eqnarray}
The determinant of $\rho_1$ is:
\begin{eqnarray}
\det \rho_1 &=& (|c_0|^2 + |c_2|^2 + |c_3|^2)|c_1|^2 - |c_0 c_1^*|^2\nonumber\\ &=& |c_1|^2 (|c_2|^2 + |c_3|^2).
\end{eqnarray}
The bipartite concurrence between subsystem $i$ and the joint subsystem $(jk)$ in a tripartite quantum system is defined as: 
\begin{eqnarray}
\mathcal{C}_{i(jk)} &=& \sqrt{2 \left( 1 - \mathrm{Tr}_i \left[ \left( \mathrm{Tr}_{jk}\, \rho \right)^2 \right] \right)},
\end{eqnarray}
where $\mathrm{Tr}_{jk} \rho$ denotes the partial trace over subsystems $j$ and $k$, yielding the reduced density matrix for subsystem $i$. The concurrence $\mathcal{C}_{i(jk)}$ quantifies the entanglement between the subsystem $i$ and the composite subsystem $(jk)$, serving as an entanglement monotone that reflects the quantum correlations based on the purity of the reduced density matrix. Using the above, the concurrence for subsystem 1 is:
\begin{eqnarray}
\mathcal{C}_{1(23)} &=& 2 |c_1| \sqrt{|c_2|^2 + |c_3|^2}\label{eq:20} \\
\mathcal{C}^2_{1(23)} &=& 4 |c_1|^2 \left( |c_2|^2 + |c_3|^2 \right).\label{eq:21}
\end{eqnarray}
Physically, $\mathcal{C}_{1(23)}$ measures the quantum correlation between resonator 1 and the combined subsystem of the atom and resonator 2. A nonzero $\mathcal{C}_{1(23)}$ indicates that resonator 1 is entangled with the joint state of subsystems 2 and 3, implying that its quantum state cannot be described independently. If $\mathcal{C}_{1(23)} = 0$, resonator 1 is unentangled, and its state is separable from subsystems 2 and 3. Similarly,
\begin{eqnarray}
\mathcal{C}_{2(31)} &=& 2 |c_2| \sqrt{|c_3|^2 + |c_1|^2}\label{eq:22} \\
\mathcal{C}^2_{2(31)} &=& 4 |c_2|^2 \left( |c_3|^2 + |c_1|^2 \right),\label{eq:23} \\
\mathcal{C}_{3(12)} &=& 2 |c_3| \sqrt{|c_1|^2 + |c_2|^2}\label{eq:24} \\
\mathcal{C}^2_{3(12)} &=& 4 |c_3|^2 \left( |c_1|^2 + |c_2|^2 \right).\label{eq:25}
\end{eqnarray}
These quantify the entanglement of subsystems 2 and 3 with their respective composite subsystems.

To determine the coefficients $c_n$, we solve the Schrödinger equation $i \frac{d|\psi\rangle}{dt} = \hat{H} |\psi\rangle$ in the steady state, using the effective Hamiltonian of Eq. (\ref{eq5}). This yields the coupled equations
\begin{eqnarray}
\left( \Delta - \frac{i \kappa}{2} \right) c_1 + J c_2 + g c_3 + \Omega_1 c_0 &=& 0, \\
\left( \Delta - \frac{i \kappa}{2} \right) c_2 + J c_1 + \Omega_2 c_0 &=& 0, \\
\left( \Delta_a - \frac{i \gamma}{2} \right) c_3 + g c_1 &=& 0,
\end{eqnarray}
where $\Delta$ is the detuning, $\kappa$ is the resonator decay rate, $\gamma$ is the atomic decay rate, $J$ is the resonator-resonator coupling, $g$ is the atom-resonator coupling, and $\Omega_1, \Omega_2$ are driving amplitudes. Defining
\begin{eqnarray}
A = \Delta - \frac{i \kappa}{2}, \quad B = \Delta - \frac{i \kappa}{2}, \quad C = \Delta_a - \frac{i \gamma}{2},
\end{eqnarray}
the equations are expressed in matrix form as
\begin{eqnarray}
M \vec{c} = \vec{d},
\end{eqnarray}
where:
\begin{eqnarray}
M =
\begin{pmatrix}
A & J & g \\
J & B & 0 \\
g & 0 & C
\end{pmatrix}, \quad
\vec{c} =
\begin{pmatrix}
c_1 \\
c_2 \\
c_3
\end{pmatrix}, \quad
\vec{d} =
\begin{pmatrix}
-\Omega_1 c_0 \\
-\Omega_2 c_0 \\
0
\end{pmatrix}.
\end{eqnarray}
Using Cramer’s rule, the coefficients are
\begin{eqnarray}
c_i = \frac{\det(M_i)}{\det(M)},
\end{eqnarray}
where $M_i$ is the matrix $M$ with its $i$-th column replaced by $\vec{d}$. The determinant of $M$ is:
\begin{eqnarray}
\det(M) &=& A B C - J^2 C - g^2 B.
\end{eqnarray}
The solutions are:
\begin{eqnarray}
c_1 &=& \frac{\left( \Delta_a - \frac{i \gamma}{2} \right) \left( \Delta - \frac{i \kappa}{2} \right) \Omega_1 - \left( \Delta_a - \frac{i \gamma}{2} \right) J \Omega_2}{\det(M)}\label{eq:34} \\ 
c_2 &=& \frac{\left( \Delta_a - \frac{i \gamma}{2} \right) \left( \Delta - \frac{i \kappa}{2} \right) \Omega_2 - \left( \Delta_a - \frac{i \gamma}{2} \right) J \Omega_1 - g^2 \Omega_2}{\det(M)}\label{eq:35} \\
c_3 &=& \frac{g \left[ -\left( \Delta - \frac{i \kappa}{2} \right) \Omega_2 + J \Omega_1 \right]}{\det(M)}\label{eq:36}
\end{eqnarray}
with:
\begin{eqnarray}
\det(M) &=& \left( \Delta - \frac{i \kappa}{2} \right)^2 \left( \Delta_a - \frac{i \gamma}{2} \right) - g^2 \left( \Delta - \frac{i \kappa}{2} \right)\nonumber\\&& - J^2 \left( \Delta_a - \frac{i \gamma}{2} \right).
\end{eqnarray}
Substituting $c_1, c_2, c_3$ into the concurrence expressions yields $\mathcal{C}_{1(23)}, \mathcal{C}_{2(31)}, \mathcal{C}_{3(12)}$, enabling the calculation of the concurrence fill as defined in prior work \textbf{[cite]}.

At resonance, we require $\det(M) = 0$. The real and imaginary parts of $\det(M)$ are:
\begin{eqnarray}
\Re(\det M) &=& \Delta_a \left( \Delta^2 - J^2 - \frac{\kappa^2}{4} \right) + \kappa \Delta \gamma - g^2 \Delta, \\
\Im(\det M) &=& -\frac{\kappa}{2} \left( \Delta^2 - \frac{\kappa^2}{4} - J^2 \right) - \kappa \Delta \frac{\Delta_a}{2} + \frac{g^2 \kappa}{2}.
\end{eqnarray}
Resonance occurs when:
\begin{eqnarray}
\Delta_a \left( \Delta^2 - J^2 - \frac{\kappa^2}{4} \right) = g^2 \Delta - \kappa \Delta \gamma.
\end{eqnarray}
The optimal atom-resonator coupling $g$ is:
\begin{eqnarray}
g = \sqrt{\frac{\Delta_a}{\Delta} \left( \Delta^2 - J^2 - \frac{\kappa^2}{4} \right) + \kappa \gamma}.
\end{eqnarray}
Similarly, the resonator-resonator coupling $J$ is:
\begin{eqnarray}
J = \sqrt{\frac{2 g^2 \Delta - \kappa \Delta \gamma}{\Delta_a} + \frac{\kappa^2}{4} - \Delta^2}.
\end{eqnarray}

Using the optimal coupling parameters $g$ and $J$ at resonance, we certify the tripartite entanglement measure, specifically the concurrence fill, in our resonator-atom-resonator model. Analytically, we compute $\mathcal{C}_{1(23)}, \mathcal{C}_{2(31)}, \mathcal{C}_{3(12)}$ using the derived coefficients $c_1, c_2, c_3$ and evaluate the concurrence fill as per the entanglement polygon inequality. Numerically, we simulate the system dynamics under the effective Hamiltonian, solving for the steady-state density matrix and computing the concurrences to validate the analytical results. A nonzero concurrence fill confirms genuine tripartite entanglement, consistent with the geometric constraints of the concurrence triangle, ensuring robust certification across all bipartitions.

\section{Results and Discussions}

Initially, if we neglect the second cavity mode ($J=0$ and $\Omega_2 = 0$), Eq. (\ref{eq2}) reduces to the simplified Jaynes-Cummings model in the rotating reference frame. In this condition, we can treat the model as a single two-level atom interacting with a single quantized cavity mode. By using this assumption, the Hamiltonian can be rewritten as
\begin{eqnarray}
\hat{H}_{JC} &=& 
\Delta_1 \hat{a}_1^{\dagger} \hat{a}_1 
+ \Delta_2 \hat{a}_2^{\dagger} \hat{a}_2 
+ \frac{\Delta_a}{2} \sigma_z + g\left( \hat{a}_1^{\dagger} \sigma_- + \hat{a}_1 \sigma_+ \right)\nonumber\\
&& 
+ \Omega_1 \left( \hat{a}_1^{\dagger} + \hat{a}_1 \right)
\label{eq29}
\end{eqnarray}
 Under this condition, the possible quantum  state takes the form
    $|\psi\rangle = C_0 |00\rangle + C_1 |01\rangle + C_2 |10\rangle$. By substituting the ansatz state $|\psi\rangle$ and into the Schrödinger equation $i\frac{d|\psi\rangle}{dt}= \hat{H}_{JC}|\psi\rangle$ evolve in a steady state, and we obtain the following solutions 
\begin{eqnarray}
\left(\Delta_1 - \frac{i \kappa}{2}\right) C_1 + g C_2 + \Omega_{1} C_0 &=& 0 \\
g C_1 + \left(\frac{\Delta_{a}}{2} - \frac{i \gamma}{2}\right) C_2 &=& 0
\end{eqnarray}
These equations can be represented in compact matrix form as
\begin{eqnarray}
\begin{pmatrix}
    \Delta_1 - \dfrac{i\kappa}{2} & g \\
    g & \dfrac{\Delta_{a}}{2} - \dfrac{i\gamma}{2}
\end{pmatrix}
\begin{pmatrix}
    C_1 \\
    C_2\\
\end{pmatrix}
& = &
\begin{pmatrix}
    -\Omega_{1} C_0 \\
    0
\end{pmatrix}
\end{eqnarray}
\begin{eqnarray}
C_1 = \dfrac{\Omega_{1} \left( \dfrac{\Delta_{a}}{2} - \dfrac{i\gamma}{2} \right)}{D}, \quad
C_2 = \dfrac{-\Omega_{1} g}{D} \label{eq:47}
\end{eqnarray}
where $D = \left(\Delta - \dfrac{i\kappa}{2}\right)
       \left(\dfrac{\Delta_{a}}{2} - \dfrac{i\gamma}{2}\right)
       - g^{2}$ represents the determinant of the system.
To get the optimal coupling parameter, we set the condition $D=0$ and obtain
\begin{eqnarray}
g &=& \sqrt{\left( \Delta - \dfrac{i\kappa}{2} \right)
\left( \dfrac{\Delta_a}{2} - \dfrac{i\gamma}{2} \right)}
\end{eqnarray}
 The density matrix, which represents the quantum state of the composite system, is 
\begin{eqnarray}
\rho &=& |\psi\rangle\langle\psi| \; = \;
\begin{pmatrix}
    |C_0|^2 + |C_2|^2 & C_0 C_1^* \\
    C_1 C_0^* & |C_1|^2
\end{pmatrix}
\end{eqnarray}
We can observe though the system has dissipation due to $\kappa$ and $\gamma$ the steady state density matrix is not diagonal. As the off-diagonal elements of the density matrix $C_1 C_0^*$ are non-zero, it contains coherence. This coherence corresponds to the quantum interference between atomic and photonic excitation of the cavity.
and we find the concurrence of this driven JC model as 
\begin{eqnarray}
\mathcal{C} &=& 2 \sqrt{ \det \rho } = 2 |C_2| |C_1|\label{eq:50} 
\end{eqnarray}
where $\det \rho = (|C_0|^2 + |C_2|^2)|C_1|^2 - |C_0|^2|C_1|^2$.
The concurrence $(\mathcal{C})$ is evaluated as a quantitative measure of atom–photon entanglement.
\begin{figure}
    \centering
    \includegraphics[width=\linewidth]{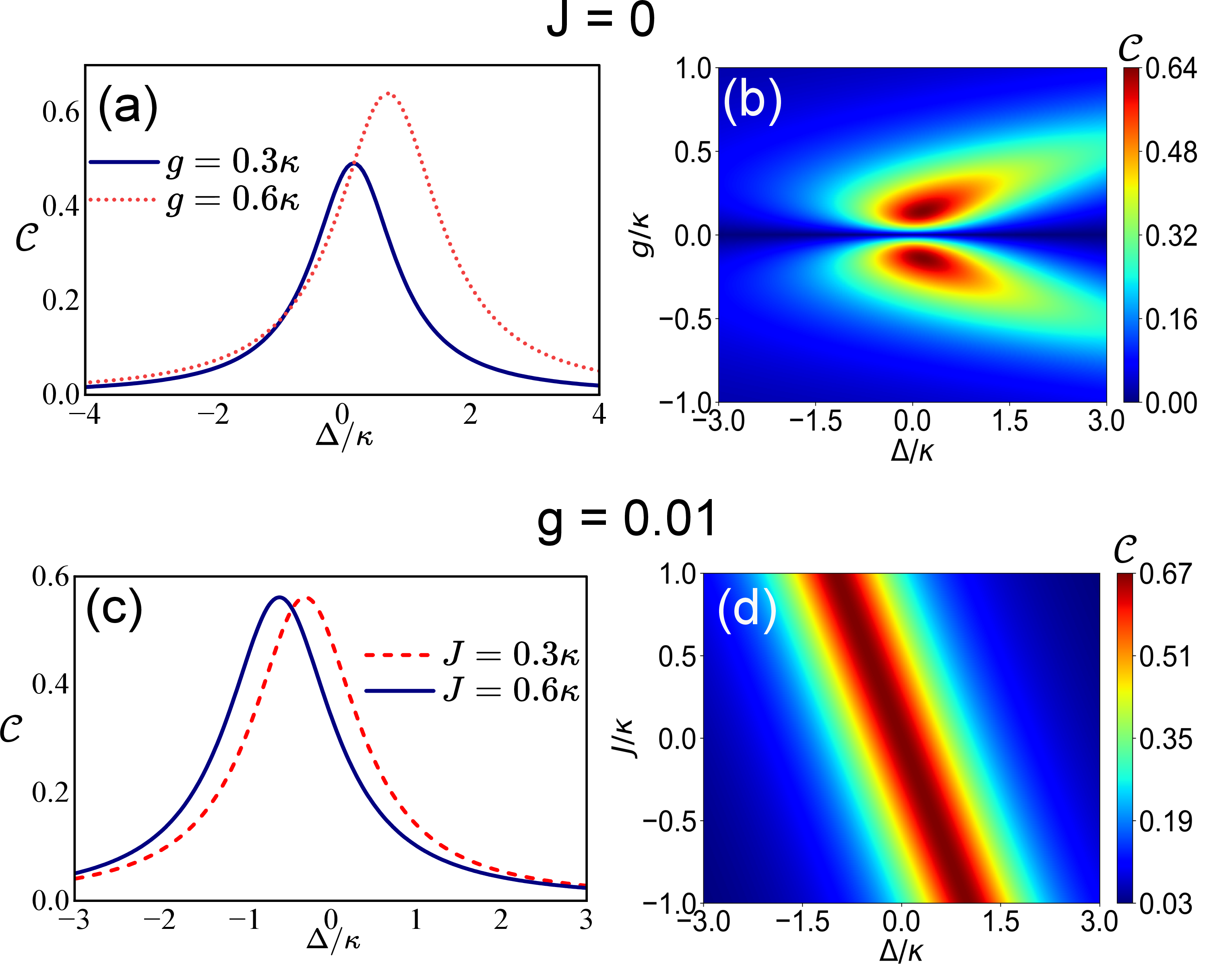}
    \caption{Variation of bipartite concurrence with normalized detuning under different coupling regimes. (a)Steady-state concurrence as a function of normalized detuning for atom–resonator coupling strengths. (b)Surface plot of concurrence as a function of normalized detuning and coupling strength. (c)Concurrence as a function of normalized detuning for two photon-hopping strengths. (d)Surface plot of concurrence as a function of normalized detuning and coupling constant J.}
    \label{Fig:FIG-2}
\end{figure}

Fig.\ref{Fig:FIG-2} illustrates the variation of bipartite concurrence between the atom and the driven cavity mode under different coupling conditions. The analysis has been carried out both analytically, within the weak‐excitation approximation, and numerically by solving the full quantum master equation in Qutip under the steady‐state condition. To generate the analytical plot, we plug the values of the coefficients $C_{1}$ and $C_{2}$ of equation \eqref{eq:47} into equation \eqref{eq:50}, and calculate the concurrence. For the plot, we have chosen experimentally feasible parameters as $\kappa=1$, $ \Delta_{a}= 1\kappa$, $ \ \gamma=0.1\kappa$, and $\Omega=0.5\kappa$

As shown in Fig.\ref{Fig:FIG-2}(a), the concurrence exhibits a distinct resonance around $\frac{\Delta}{\kappa}\approx 0$ where the cavity and the atomic transitions are nearly resonant. This corresponds to the maximum hybridization between the atomic excitation $|e,0\rangle$ and the single photon state $|g,1\rangle$. The peak of the concurrence reflects the coherent exchange of excitation between two subsystems. 
 
The term $g\left( \hat{a}_1^{\dagger} \sigma_- + \hat{a}_1 \sigma_+ \right)$ couples the states 
$|e,0\rangle \leftrightarrow |g,1\rangle$ with coupling strength g. For larger coupling strength $g=0.6\kappa$ the splitting between dressed states increases and the concurrence peak broadens due to enhanced atom-field mixing. For weak coupling $g=0.3\kappa$ the hybridization is reduced resulting in a narrower and lower concurrence peak. These features clearly signify the role of the atom-resonator coupling in generating and controlling bipartite entanglement.

From Fig.\ref{Fig:FIG-2} (b) it can be observed that two bright lobes appear symmetrically about resonance, corresponding to the dressed state splitting. The concurrence reaches its maximum when the energy exchange between atom and cavity is stronger. For large detuning, the interaction becomes off-resonant, which suppresses atom-photon coherence, and the concurrence is zero. 

  Fig.2(c) and(d) illustrate the bipartite concurrence when the photon–hopping interaction between the two resonators is happening at very weak($g=0.001\kappa$) atom-resonator coupling. Under this condition, atom–field coupling becomes negligible, and the dominant correlation arises from photon delocalization between the two coupled cavities.Fig.2.(c) shows that increasing J from $0.3\kappa$ to $0.6\kappa$ broadens and slightly shifts the concurrence peak, as the normal modes of the coupled resonators hybridize. The corresponding surface plot in Fig.2(d) confirms that the concurrence now scales mainly with $\frac{J}{\kappa}$, showing near‐linear dependence in the weak‐coupling limit.
Physically, this indicates that entanglement is transferred from atom–cavity correlations to inter-cavity photonic correlations as the hopping interaction becomes dominant.
\begin{figure}
    \centering
    \includegraphics[width=\linewidth]{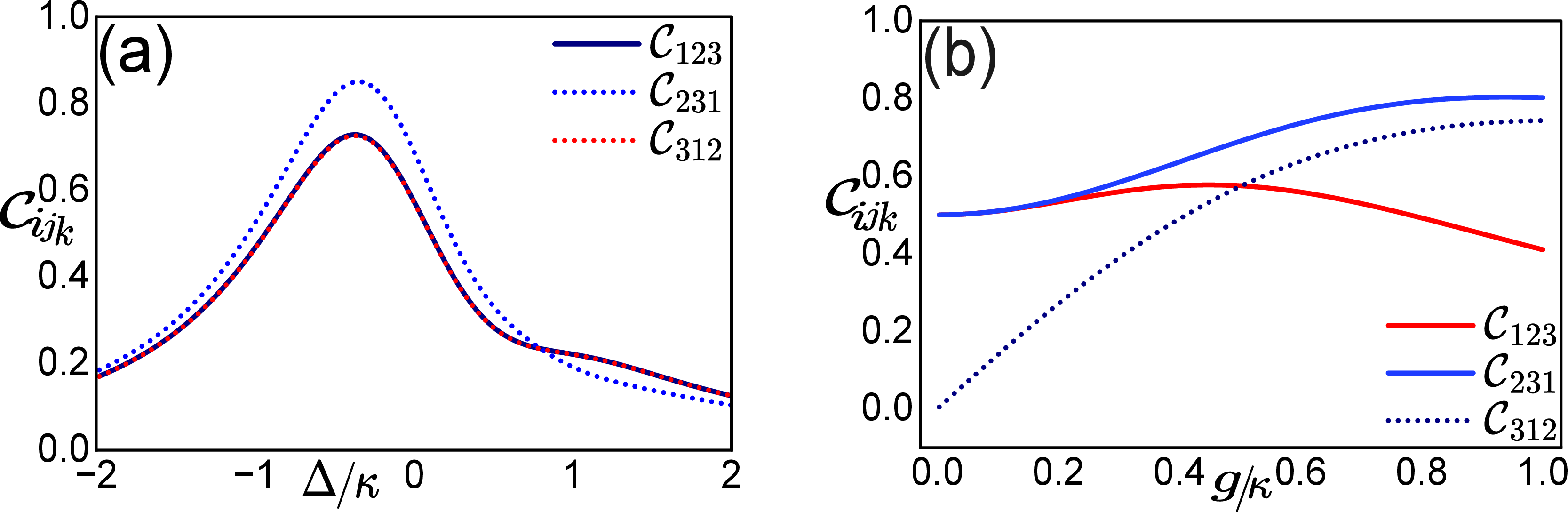}
    \caption{(a) Variation of the bipartite concurrences as a function of normalized detuning in the hybrid regime where both the atom–cavity coupling and the inter-cavity hopping are finite. (b)Dependence of the three concurrences on the normalized atom–cavity coupling at fixed detuning. }
    \label{fig:FIG-3}
\end{figure}
\begin{figure}
    \centering
    \includegraphics[width=\linewidth]{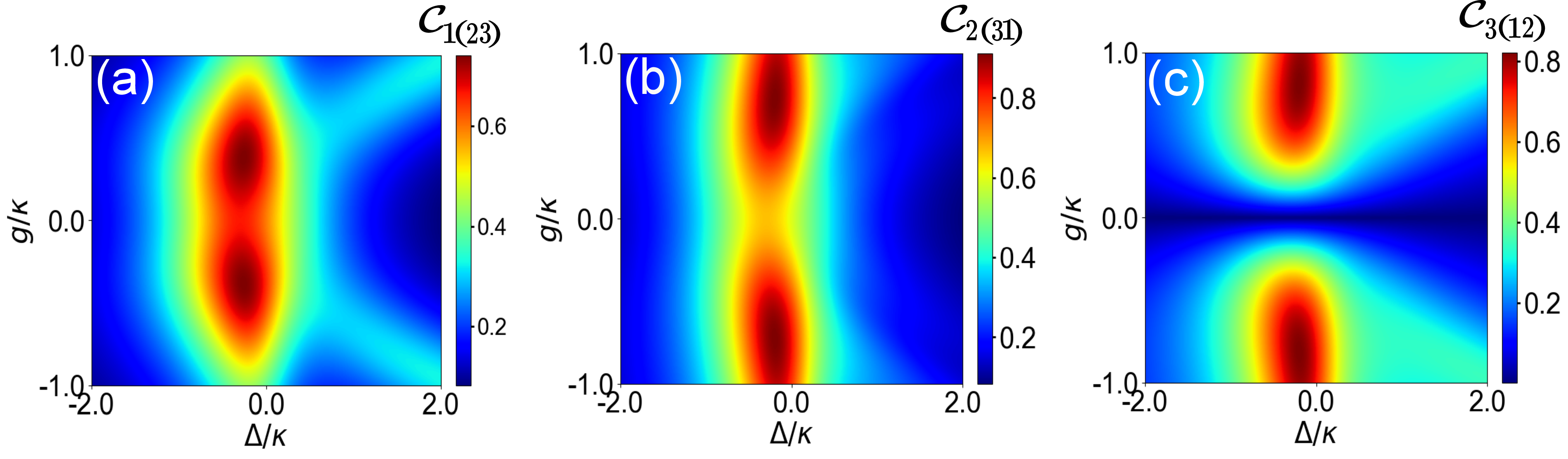}
    \caption{Surface plots of bipartite concurrences as functions of normalized detuning and atom–cavity coupling strength g.}
    \label{fig:6}
\end{figure}

Next, we analyze the tripartite entanglement when both the couplings are present (finite $g$ and $J$). 
Initially, we consider the atom to be in the
superposition state $\frac{1}{\sqrt{2}}(|0\rangle+|1\rangle)$ while both the resonator is in the vacuum state $|0\rangle$.We use a  conditional photon
exchange interactions by Jaynes–Cummings interaction to
entangle the atom with the first resonator directly. Through
controlled time evolution, the excitation is coherently
transferred to the second resonator, such that when the atom
is excited, and both resonators are also excited simultaneously.
After symmetric photon exchange by controlling each
distinct photon excitation paths with corresponding
probabilities computed from the Schrödinger equation and
interaction blocking (via quantum interference), the final
state collapses to an entangled state. By applying symmetric, weak coupling between the atom and both resonators, we allow the single excitation to coherently distribute among all three subsystems. With precisely tuned coupling strengths and interaction times, such that destructive interference blocks multi-excitation pathways, while constructive interference amplifies the single-excitation symmetric state. The system evolves into a coherent superposition where exactly one of three subsystems holds the excitation, forming a robust entangled state.

\begin{figure}
    \centering
\includegraphics[width=\linewidth]{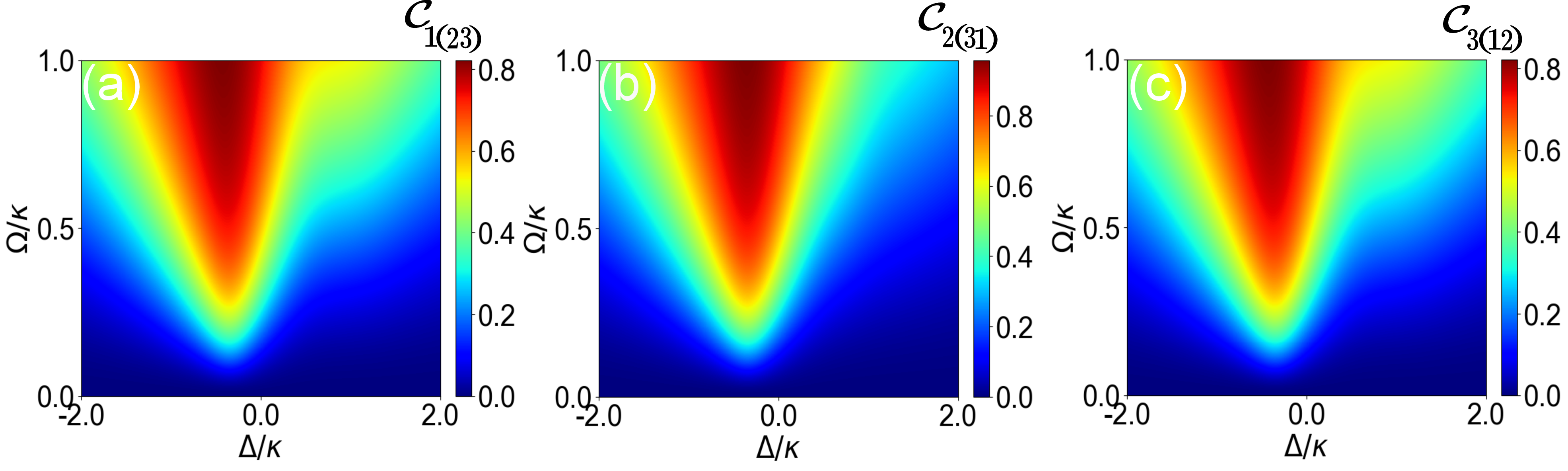}
    \caption{Surface plots of bipartite concurrences as functions of normalized detuning and drive strength for nonzero atom–cavity (g) and inter-cavity (J) couplings. The bright regions mark the optimal drive, where coherent excitation exchange among the atom and both cavities generates maximal hybrid entanglement.}
    \label{fig:FIG-7}
\end{figure}

When both the atom–cavity coupling (g) and the inter-cavity hopping (J) are nonzero, the system enters a hybrid tripartite regime where atomic and photonic degrees of freedom are coherently mixed.Fig.3 illustrates the evolution of the three bipartite concurrences $\mathcal{C}_{1(23)}, \mathbf{C}_{2(31)}$ and $\mathcal{C}_{3(12)}$, which correspond to the sides of the concurrence triangle used to quantify the distribution of entanglement among the atom and the two resonators. We have analytically calculated $\mathcal{C}_{1(23)}, \mathbf{C}_{2(31)}$ and $\mathcal{C}_{3(12)}$ by plugging the coefficients of $C_{1},C_{2}$ and $C_{3}$ from the equation \eqref{eq:34},\eqref{eq:35} and \eqref{eq:36} into the equation \eqref{eq:20},\eqref{eq:22} and \eqref{eq:24}.Fig.2(a) illustrates the concurrence peaks near the resonance condition $\Delta\approx0$, where the hybridized normal modes of the coupled system become degenerate and the excitation is maximally shared among the three subsystems. The near overlap of $C_{1(23)}$ and $C_{2(31)}$arises from the structural symmetry between the two resonators, while the smaller magnitude of $C_{3(12)}$indicates that atomic participation is slightly suppressed due to spontaneous emission and detuning from the photonic resonance. Fig.\ref{fig:6}(b) shows the dependence of the concurrences on the normalized coupling strength. For weak coupling, the system remains predominantly photonic, and the entanglement is mainly confined between the two resonators via photon hopping J. As g increases, atom–photon coherence strengthens, leading to an enhanced redistribution of correlations among all three systems. In the strong-coupling regime, the three bipartite concurrences become comparable, signaling the formation of genuine tripartite hybrid entanglement where no single subsystem can be factorized from the others. This transition reflects the eventual balance between atom-mediated nonlinearity and photon hopping coherence, establishing a steady-state hybridized field that is a coherent superposition of atomic and photonic excitations.
\begin{figure}
    \centering
    \includegraphics[width=\linewidth]{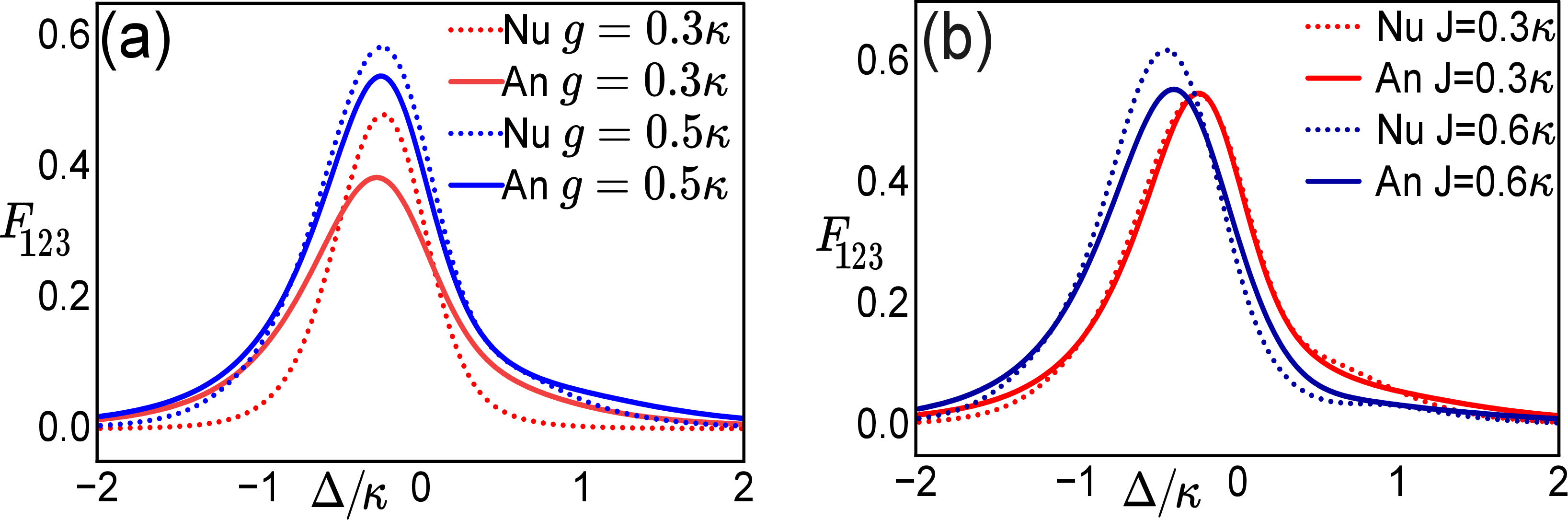}
    \caption{Comparison of analytical (solid) and numerical (dotted) results for the tripartite entanglement measure(concurrence fill) as a function of normalized detuning.(a) Variation with atom–cavity coupling g at fixed J. Stronger light–matter coupling enhances the hybridization between atomic and photonic excitations, yielding higher peak entanglement near resonance.(b) Variation with inter-cavity coupling J at fixed g. Increasing photon hopping broadens the resonance region and facilitates coherent excitation sharing across the full tripartite system.}
    \label{fig:FIG-5}
\end{figure}

Fig.4 shows the behavior of the bipartite concurrence terms $\mathcal{C}_{1(23)}, \mathbf{C}_{2(31)}$ and $\mathcal{C}_{3(12)}$, which quantify how each subsystem (resonator 1, resonator 2, and the atom) is entangled with the remaining two subsystems as a function of the normalized detuning and coupling strength. In fig.4(a) $\mathcal{C}_{1(23)}$ exhibits two distinct resonance lobes symmetrically placed around $\Delta=0$, reflecting the hybridization between the first resonator and the collective atom–photon excitation manifold. This splitting corresponds to the well-known vacuum Rabi splitting in cavity QED, where the resonance condition is modified by the coupling strength g. As g increases, these lobes move apart, indicating stronger mixing between the photonic and atomic components in the normal-mode spectrum.
Fig.5(b) $\mathcal{C}_{2(31)}$ displays a single pronounced vertical ridge near $\Delta=0$.This suggests that the second resonator becomes maximally entangled with the atom–resonator-1 subsystem at resonance, where coherent photon exchange between the resonators (mediated by J) facilitates delocalization of the excitation across the network. In contrast,fig.5(c) $\mathcal{C}_{3(12)}$ shows a complementary structure, with the entanglement peaking for small coupling strength and decreasing as g increases. Physically, this indicates that at weak coupling the atom shares coherence with both resonators, but as g\ becomes large, the atom–resonator-1 subsystem becomes more isolated from resonator 2, reducing the atom’s joint participation. All these plots reveal that the three bipartite concurrences are not independent but redistribute among the subsystems as the coupling strength and detuning are tuned.

From \ref{fig:FIG-7} (a) Shows as $\Omega$ increases from zero, the drive pumps photons into cavity 1, which couples both to the atom (through g) and to cavity 2 (through J).For small $\Omega$, entanglement is weak because the field has insufficient amplitude to establish correlations between the subsystems. Around an optimal intermediate $\Omega$ (visible as the bright region at moderate $\frac{\Omega}{\kappa}\approx 0.4-0.6$, $\mathcal{C}_{1(23)}$  peak indicates maximum hybridization of the cavity 1 mode with both atom and cavity 2.  \ref{fig:FIG-7} (b) Its profile is almost similar but slightly shifted because cavity 2 couples indirectly to the atom via photon hopping J. Increasing $\Omega$ enhances the photonic population transfer through J, producing strong photon–photon entanglement even slightly off resonance. The maximum again appears at moderate, $\Omega,$ showing optimal photon delocalization between the two cavities.\ref{fig:FIG-7} (c) illustrates that $\mathcal{C}_{3(12)}$ rises sharply with increasing $\Omega$ and saturates at high $\Omega$, implying that the atomic excitation becomes strongly dressed by the photonic fields. The red region near zero detuning corresponds to the resonance condition where the dressed atomic–photonic state is maximally entangled.
The bright regions mark the optimal drive, where coherent excitation exchange among the atom and both cavities generates maximal hybrid entanglement.
  
\begin{figure}
    \centering
    \includegraphics[width=\linewidth]{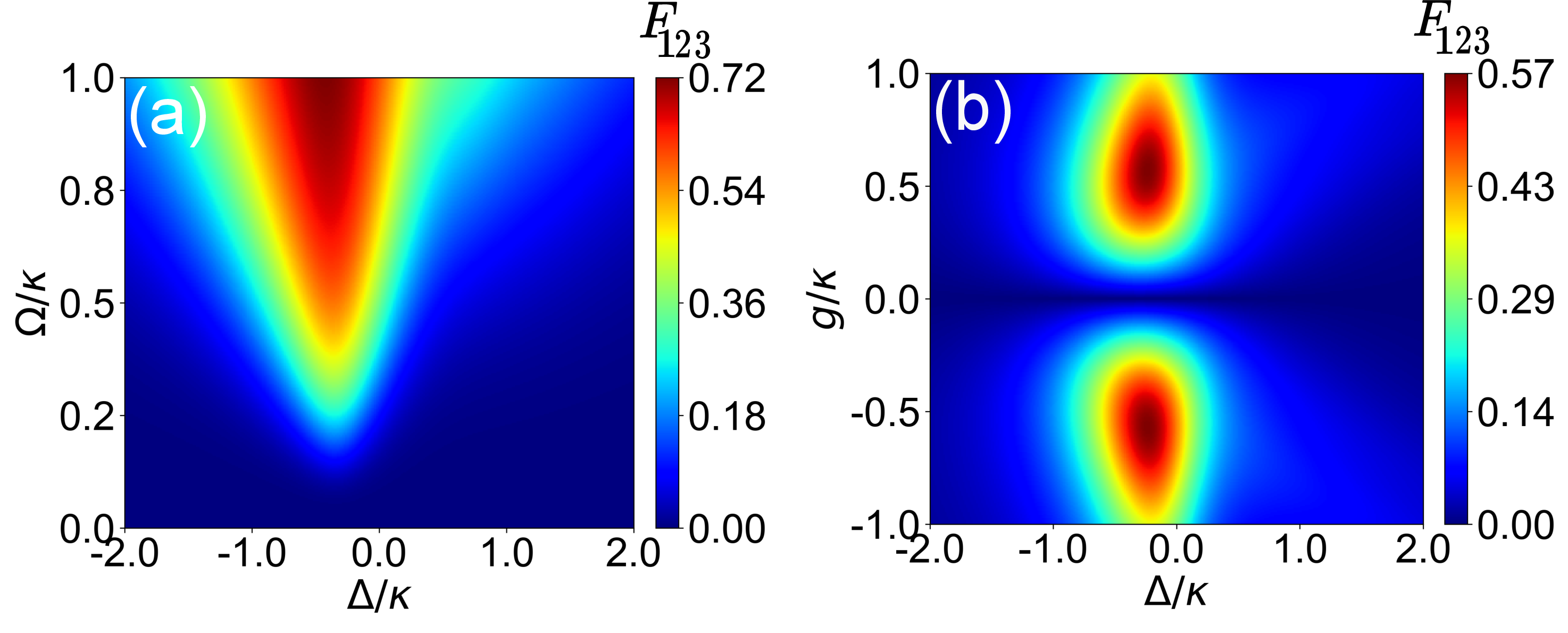}
    \caption{Color plots of the tripartite entanglement measure as functions of normalized detuning.(a)Dependence on driving strength showing drive-induced enhancement of hybrid atom–photon correlations near resonance. (b)Dependence on atom–cavity coupling where maximal entanglement occurs at resonance. }
    \label{fig:FIG-4}
\end{figure}

To analytically calculate the concurrence fill$(\mathcal{F}_{123})$, measure of tripartite entanglement, we first calculate $\mathcal{C}^2_{1(23)}, \mathcal{C}^2_{2(31)}$ and $\mathcal{C}^2_{3(12)}$ by plugging the values of $c_{1},c_{2}$ and $c_{3}$ from equations \eqref{eq:34},(35), and \eqref{eq:36} using the expressions of equation \eqref{eq:21},\eqref{eq:23} and \eqref{eq:25}.Then, by putting the values of $\mathcal{C}^2_{1(23)}, \mathcal{C}^2_{2(31)}$ and $\mathcal{C}^2_{3(12)}$ in equation\eqref{eq:8} we have calculated concurrence fill. For the numerical plot, we solved the full quantum master equation in the steady state regime using Qutip. We have used experimentally feasible parameters for the numerical plot, such as $\kappa=1, \gamma=0.1\kappa, \Delta_{a}=i\kappa,\Omega=0.5\kappa$ $J=0.5\kappa$(For fixed J),$g=0.5\kappa$(For fixed g).
Fig.6(a) illustrates the variation of concurrence fill($\mathcal{F}_{123}$) with the atom-resonator coupling strength with fixed inter resonator coupling.The atom couples more strongly to cavity 1, enhancing the exchange of excitations between the atom and the photon field. This strong light–matter coupling generates a greater degree of hybridization between the atomic and photonic states, producing higher tripartite correlations that extend to cavity 2 through photon hopping. Consequently, the peak of $\mathcal{F}_{123}$ becomes higher and narrower near resonance signifying a stronger and more coherent entanglement among the three subsystems.The enhancement with g arises because increasing the atom–photon coupling increases the dressed-state splitting and improves the coherent excitation transfer that generates entanglement across the entire hybrid system. Fig.6(b) shows the variation of $\mathcal{F}_{123}$ with the inter resonator coupling strength for fixed atom- resonator coupling.The two resonators exchange photons via J, forming delocalized photonic modes. A moderate J allows the excitation initially coupled to the atom through cavity 1 to propagate coherently into cavity 2, thereby building up genuine tripartite entanglement. As J increases from $0.3\kappa$  to $0.6\kappa$ the entanglement peak broadens slightly—indicating that stronger photon tunneling promotes energy-exchange coherence across the entire system, even slightly off resonance. The excellent agreement between numerical and analytical curves confirms the validity of the weak-drive analytical model. 

\begin{figure}
    \centering
    \includegraphics[width=\linewidth]{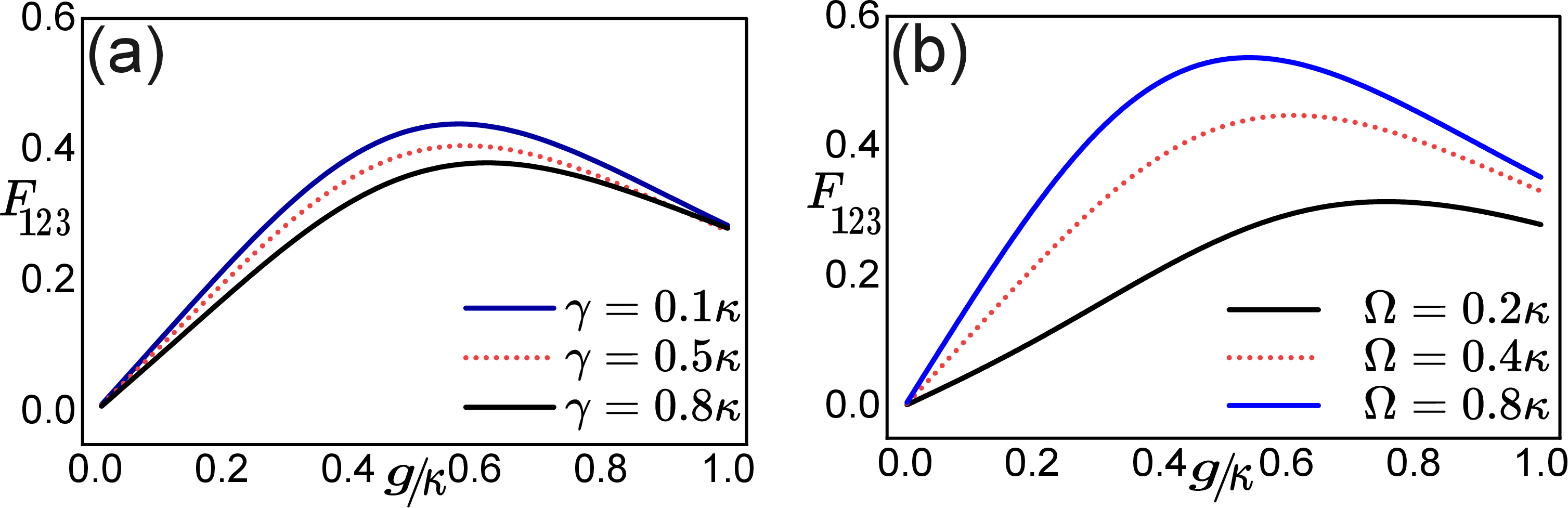}
    \caption{Variation of tripartite entanglement with normalized atom–resonator coupling.(a) Dependence on the atomic decay rate shows suppression of hybrid correlations with increasing dissipation.(b) Dependence on driving strength with entanglement.}
    \label{fig:FIG-8}
\end{figure}
 The driving strength controls the number of coherent excitations injected into the system. From Fig.7(a) we can observe For small $\Omega$,the excitation probability is weak, so entanglement remains minimal. As $\Omega$ increases, the coherent population of the resonators enhances correlation between atomic and photonic degrees of freedom, resulting in a strong rise in $\mathcal{F}_{123}$. The V-shaped region centered at $\Delta\approx 0$ reflects the resonance condition where the driving field frequency matches the coupled system’s dressed-state energies, maximizing coherent energy exchange.  This indicates that the tripartite entanglement is drive-assisted, the external field creates correlated excitations among the atom and both resonators most efficiently when driven near resonance.
 Fig.7 (b) illustrates how increasing g strengthens hybrid light–matter coupling, therefore amplifying the system’s collective quantum correlations and enabling genuine tripartite entanglement. At $g\approx 0$, the system reduces to two weakly coupled photonic modes, giving almost zero $\mathcal{F}_{123}$.As g increases, energy hybridization between atomic and photonic excitations generates strong tripartite correlations, seen as the bright central region. The symmetric lobes around $\Delta=0 $ correspond to the upper and lower polaritonic branches of the Jaynes–Cummings ladder, where coherent superpositions of atom–photon states form. The entanglement is maximal at these hybridized resonances, where the atom effectively mediates non-classical correlations between the two cavities. 

Fig.8 illustrates how the tripartite entanglement measure $\mathcal{F}_{123}$ varies with the normalized atom–resonator coupling strength for different atomic decay rates ($\gamma$) and driving strengths($\Omega$) which provides a clear picture of how dissipation and external drive jointly influence multipartite hybrid entanglement in the system. Fig.8(a) Shows as the coupling g increases from zero, the system transitions from a weak-coupling to a strong-coupling regime, enhancing coherent energy exchange between the atom and resonator 1.$\mathcal{F}_{123}$ increases initially, reaching a maximum at an optimal $\frac{g}{\kappa}$ then decreases as excessive coupling splits the resonance and leads to reduced population overlap among the three modes.Increasing $\gamma$(atomic decay) suppresses the entanglement amplitude since stronger spontaneous emission destroys atomic coherence and weakens atom–photon correlations.The observed peak reflects the balance between coherent atom–photon interaction and dissipative losses, characteristic of the photon-blockade-assisted hybrid entanglement regime. Small 
$\gamma$ allows the atom to participate in collective hybridization, while large $\gamma$ rapidly projects the atom to its ground state, thereby washing out tripartite correlations.Fig.8(b) depicts the drive field acts as a tunable resource for injecting correlations into the coupled resonator–atom network. For fixed dissipation, increasing the external drive populates the resonators more efficiently, facilitating enhanced atom–photon and inter-cavity correlations, hence a rise in $\mathcal{F}_{123}$.The behavior shows a nonlinear response where entanglement is maximized under moderate coherent driving—strong enough to induce hybridization but weak enough to avoid saturation and decoherence.

\section{Conclusion}
We have analyzed a hybrid atom coupled dual micro-resonator system exhibiting controllable bipartite and tripartite entanglement in the weak-driving regime.. Analytical results derived from steady-state solutions show excellent agreement with those from full numerical simulations. When only the atom–cavity coupling(g)is present, strong atom–photon entanglement emerges via the Jaynes–Cummings interaction. Introducing photon hopping couples the two resonators, extending the correlations into a genuine tripartite regime. The concurrence-based measures reveal how the interplay between and detuning governs the transition from localized to delocalized hybrid entanglement. These results demonstrate that engineered coupling and drive parameters can coherently control multipartite quantum correlations, paving the way for tunable entanglement generation in cavity-QED networks.
\bibliographystyle{unsrt}
\bibliography{my}
\end{document}